\documentclass[conference]{IEEEtran}
\makeatletter
\makeatletter
\usepackage{amsmath,amssymb}
\usepackage{graphicx}
\usepackage{cite}
\usepackage{multirow, booktabs}
\usepackage{array}
\usepackage{mathtools}
\usepackage{tabularx}
\usepackage{siunitx}
\usepackage[nolist]{acronym}
\usepackage{subfig}
\usepackage{acronym}
\usepackage{stmaryrd}
\usepackage{color}
\usepackage{xfrac}
\usepackage{soul}

\usepackage[T1]{fontenc}
\usepackage[utf8]{inputenc}

\renewcommand{\baselinestretch}{0.988}

\begin{document}
 \makeatletter
  \addtolength{\textheight}{-0.1in}  
 \makeatother
\title{\vspace{0cm}3GPP-Compliant Radar Cross Section Characterization of Indoor Factory Targets\vspace{0cm}}

\author{
\IEEEauthorblockN{Ali Waqar Azim\IEEEauthorrefmark{1}, Ahmad Bazzi\IEEEauthorrefmark{1}\IEEEauthorrefmark{2},Roberto Bomfin\IEEEauthorrefmark{1}, Marwa Chafii\IEEEauthorrefmark{1}\IEEEauthorrefmark{2}}
\IEEEauthorblockA{\IEEEauthorrefmark{1}Engineering Division, New York University Abu Dhabi (NYUAD), UAE}
\IEEEauthorblockA{\IEEEauthorrefmark{2}NYU WIRELESS, Tandon School of Engineering, New York University, Brooklyn, NY, USA}
\IEEEauthorblockA{Emails: \{ali.waqar.azim,ahmad.bazzi,roberto.bomfin,marwa.chafii\}@nyu.edu}
}

\maketitle

\begin{abstract}
The following paper presents a systematic 3rd Generation Partnership Project (3GPP)-compliant characterization of radar cross section (RCS) for indoor factory (InF) objects, including small and mid-sized unmanned aerial vehicles (UAVs), robotic arms, and automated guided vehicles (AGVs). Through measurements in the \SIrange{25}{28}{\GHz} range, we validate the 3GPP standardized log-normal distribution model for RCS for above-mentioned target objects. The 3GPP-complaint RCS parameters obtained for the small-sized UAV are in close agreement (\(<1\) dB deviation) with 3GPP agreed values. The mid-sized UAVs exhibit higher reflectivity compared to the small-sized UAV due to enhanced specular components attributed to material and lithium-ion battery packs. The robotic arm exhibits dynamic RCS behavior due to mechanical articulation, whereas UAVs show clear size-dependent reflectivity patterns in AGVs. Our findings provide empirical validation for RCS characterization for integrated sensing and communication channel modeling in InF environments.
\end{abstract}
\section{Introduction}
Integrated sensing and communication (ISAC) has emerged as one of the key vertical for upcoming 6G networks \cite{10901856}, promising to unify environmental perception and data transmission within shared infrastructure and spectral resources. ISAC channel modeling requires complete characterization of the propagation conditions and the objects present within the environment, where each object has a unique signature, which can be characterized by its radar cross section (RCS). The RCS is an important statistic for future ISAC designs, especially for multi-static target detection and power allocation \cite{10494224}.

The standardized 3rd Generation Partnership Project (3GPP) TR38.901 \cite{3gpp38901} channel model lacks ISAC-related components, including explicit RCS characterization of objects. Furthermore, it employs an oversimplified scattering paradigm that treats all objects as passive reflectors. Thus, these models fail to account for the target-specific RCS variations that are fundamental for sensing applications. To address this limitation, the 3GPP RAN1 working group has initiated the development of ISAC channel model, which will explicitly incorporate RCS-based target definitions within its framework. In the context of RCS characterization, a three parameter-based RCS modeling approach has been agreed upon, where the first parameter defines the mean RCS, the second captures angle-dependent contributions, and the third models RCS variations using a log-normal statistical distribution. In addition, the RAN1 working group has provided guidelines on how to model the three parameters for several objects, including small-sized and large-sized unmanned aerial vehicles (UAVs), humans, and automotive vehicles. 

In 3GPP's RCS characterization approach, the degree of angular dependency varies across object classes. For instance, simplified models omitting angle dependence are employed for small-sized UAVs and humans, whereas RCS profiles requiring precise angle-dependent modeling are necessary for automotive vehicles and large-sized UAVs. However, other target objects commonly present in an indoor factory (InF) environment, such as mid-sized UAVs, robotic arms, and automated guided vehicles (AGVs), have not yet received explicit modeling guidance from the 3GPP. Although the agreed-upon three-parameter RCS model applies, the current standardization lacks specificity regarding the modeling of the angular component for these targets.

Recent advances in UAV sensing have yielded extensive work on RCS measurement and statistical modeling across various frequencies and platforms. \cite{ezuma2021radar} proposed a statistical recognition framework using RCS measurements at \SI{15} and \SI{25}{\GHz}. Complementing this, \cite{rosamilia2022radar} provided RCS measurements from \SIrange{8.2}{18}{\GHz}, while \cite{ezuma2022comparative} evaluated machine learning and deep learning algorithms for RCS-based UAV classification. The survey by \cite{patel2018review} outlines key radar classification and RCS modeling approaches for small UAVs. In the ISAC context, \cite{azim2025statistical} reported monostatic and bistatic RCS measurements of UAVs, robotic arms, and AGVs at \SIrange{25}{28}{\GHz}. Moreover, \cite{semkin2021drone} assessed RCS-based drone detection using \SI{28}{\GHz} measurements, and \cite{semkin2020analyzing} analyzed RCS of UAVs across \SIrange{26}{40}{\GHz}. 
Existing literature still lacks RCS measurements for industrial assets such as robotic arms and AGVs. Researchers have yet to propose a standardized ISAC channel model that incorporates the three RCS parameters specified by 3GPP.

To provide a practical perspective, this paper presents a 3GPP-compliant RCS characterization methodology for InF targets, addressing three key challenges: (1) quantifying log-normal RCS fluctuations across target objects, (2) validating small-sized UAV parameters against 3GPP standards, and (3) establishing baseline RCS models for InF equipment.

The remainder of this paper is organized as follows: Section~\ref{sec2} details the 3GPP RCS standardization framework. Section~\ref{sec3} presents our RCS characterization methodology. Section~\ref{sec4} analyzes the experimental results, validating log-normal distribution fits 3GPP standard-based parameters and summarizes key findings regarding RCS characteristics of small/mid-sized UAVs, robotic arms, and AGVs. Finally, Section~\ref{sec5} draws conclusions.
\section{Standardized RCS Characterization Framework by 3GPP}\label{sec2}
The 3GPP RAN\#116bis  agreement establishes that the standardized RCS characterization framework accounts for multiple deterministic and stochastic dependencies of the scattering properties of a physical object including object's physical dimensions, material parameters, geometric configurations, orientation relative to the transceiver, the incident and scattered angles, and the operating frequency. Subsequent RAN\#118bis agreement formalizes two models for RCS characterization. The first model implements RCS as multiplicative decomposition, i.e., \(\mathrm{RCS} = A \times B\) where \(A\) is the mean RCS value invariant to angle variations, and \( B \) captures stochastic fluctuations through a unit-mean log-normal distribution. The mean RCS term $A$ serves as the primary parameter of object-specific properties and incorporates size/material/geometry effects and handles frequency scaling. The second model extends RCS modeling by explicitly incorporating an angular dependency term, resulting in \(\mathrm{RCS} = A \times B_1 \times B_2\), where \( B_1\) is a deterministic angle-dependent function, and \( B_2 \) is generated by unit-mean log-normal distribution. 

For UAVs with single scattering points in monostatic configurations, RAN\#118bis specifically selects the three-component model, i.e., \(\mathrm{RCS} = A \times B_1 \times B_2\) with distinct implementations based on size class. Small-sized UAVs utilize $B_1 = 1$, i.e., \(0\) dB with $A$ as the mean RCS value, while large-sized UAVs incorporate angle-dependent $B_1$ components while maintaining $A$ as the mean RCS. In both cases, $B_2$ follows the standardized unit-mean log-normal distribution. Building upon these foundations, RAN\#119bis specifies that for humans modeled as single scattering points in monostatic configurations, two RCS models are supported. The first model sets $B_1 = 0$ dB with $A$ as mean RCS and $B_2$ following log-normal distribution, while the second model incorporates angle-dependent $B_1$ implemented either as: (i) antenna radiation pattern per TR38.901, (ii) analytic function, or (iii) lookup table, with $B_2$ maintaining the unit-mean log-normal characteristics. For vehicles with single scattering points, the framework mandates an explicit modeling for angle-dependent $B_1$ with the same implementation options, while multi-scatter point vehicle models prescribe five specific scattering locations (front, left, back, right, roof) each following the $\mathrm{RCS} = A \times B_1 \times B_2$ formulation. The standard constrains $B_2$ through upper bounding $k\sigma$ dB where $\sigma$ is the standard deviation of $B_2$ in dB, with $k=3$ subsequently adopted in RAN\#120bis. 

The RAN\#120bis agreements further refine bistatic and monostatic consistency requirements, mandating that bistatic RCS values must align with monostatic results when incident and scattered angles coincide. Moreover, the 3GPP RAN\#120bis agreement has established standardized RCS parameters through its consensus-driven process, yielding definitive values for both small-sized UAVs and humans. For UAV characterization, the agreed parameters comprise $A = -12.81$ dBsm, $B_1 = 0$ dB, and $B_2 = 3.74$ dB. The identical three-parameter model extends to human targets with values $A = -1.37$ dBsm, $B_1 = 0$ dB, and $B_2 = 3.94$ dB. 
The standardized parameters originate from the 3GPP evaluation framework, which forms a standard format of the measurement data submitted by all participating members through systematic averaging.
%
\section{RCS Characterization}\label{sec3}
\subsection{RCS Characterization Methodology}
In our RCS characterization approach, we follow the three-component model $\mathrm{RCS} = A \times B_1 \times B_2$ with modeling the angle-dependent component, $B_1 = 0$ dB for mid-sized UAVs, robotic arms, and AGVs. The 3GPP standardization maintains uniformity in the treatment of $A$ and $B_2$ across all object classes, including humans, UAVs, and vehicles, with the primary distinction arising only in $B_1$. Although 3GPP explicitly defines angle-dependent $B_1$ for large-sized UAVs and vehicles, it does not provide guidance for mid-sized UAVs, robotic arms, and AGVs. Our RCS characterization methodology assumes $B_1 = 0$ dB as: (1) it aligns with the 3GPP’s treatment of target objects with negligible angle dependence (e.g. small-sized UAVs/humans), and (2) for our dynamic experimental conditions, angular effects are inherently embedded in the measurements as the UAV platforms undergo continuous flight rotations, while robotic arms and AGVs execute typical InF task-driven movements, ensuring angular diversity in the captured RCS statistics. Thus, the components \(A\) and $B_2$ describe the variations through the log-normal distribution, which provides a conservative and measurement-oriented baseline consistent with 3GPP’s framework.
\subsection{Measurement Setup and Considerations}
 \begin{figure}[t]
     \centering
     \includegraphics[trim={2mm 0mm 1mm 0mm},clip, width=1\linewidth]{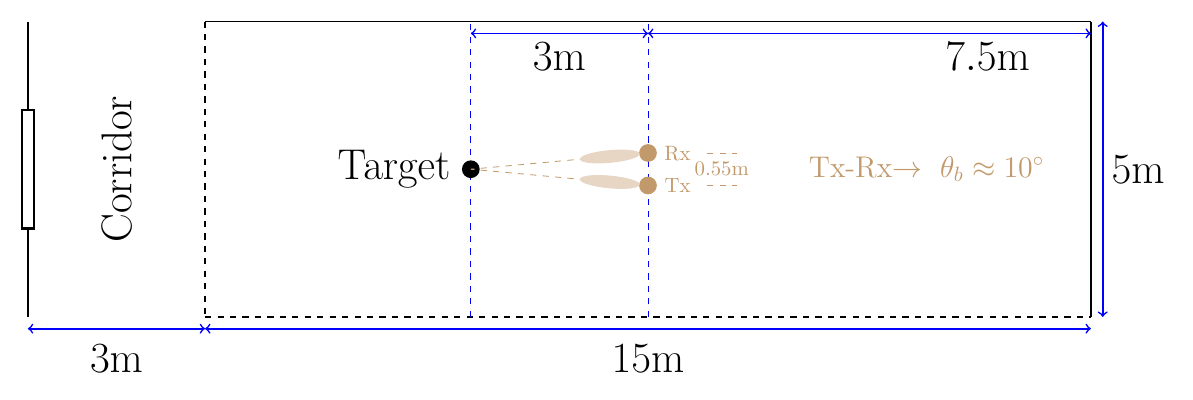} 
   \caption{\textcolor{black}{The layout of the measurement environment.}}
     \label{layout} 
 \end{figure}
The measurement campaign was conducted in the KINESIS Lab, Core Technology Platforms, at NYU Abu Dhabi, a dedicated \SI{5}{\meter} (width) $\times$ \SI{15}{\meter} (length) $\times$ \SI{8.5}{\meter} (height) indoor facility designed to emulate InF environments. The lab features characteristic InF architectural elements including elevated ceilings and open spatial configurations. The experimental layout is shown in Fig.~\ref{layout}.
 \begin{figure}[t]
     \centering
     \includegraphics[trim={2mm 0mm 1mm 0mm},clip, width=1\linewidth]{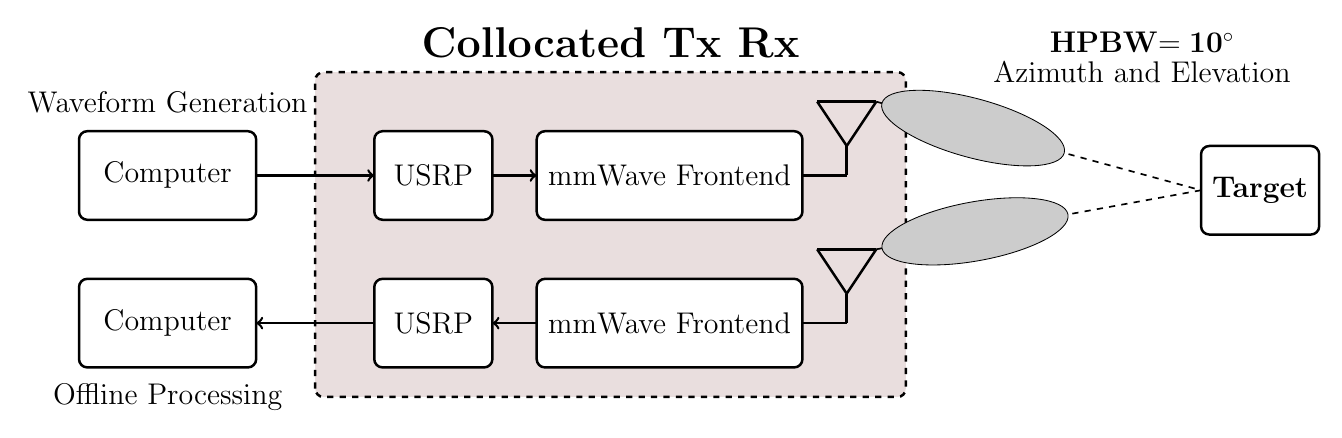} 
    \caption{\textcolor{black}{Measurement testbed configuration.} }
     \label{testbed} 
 \end{figure}
\begin{table}[tb]
\centering
\caption{Key System Parameters}
\label{tab:system_parameters}
\begin{tabular}{|l|l|}
\hline
\multicolumn{2}{|c|}{\textbf{Measurement Setup Parameters}} \\
\hline
Operating Frequency & \SIrange{25}{28}{\GHz} \\
\hline
Intermediate Frequency & \SI{3}{\GHz} \\
\hline
RFIC & Sivers EVK02004 \\
\hline
HPBW & \(10^\circ\) (Azimuth and Elevation) \\
\hline
RFIC ADC & \(10\) bits \\
\hline
USRP ADC & \(12\) bits \\
\hline
Polarization & H-H \\
\hline
Measurement Conditions & LoS\\
\hline
USRP & B205mini \\
\hline
USRP Bandwidth & \(20\,\mathrm{MHz}\) \\
\hline
Signaling & Zadoff-Chu \\
\hline
Tx-Rx Height & \(1\) m from ground \\
\hline
Dataset Size & \(300\) MB \\
\hline
Location & NYU Abu Dhabi \\
\hline
\end{tabular}
\end{table}

The monostatic measurement testbed is shown in Fig. \ref{testbed} which employs a collocated transmitter (Tx)/receiver(Rx). The measurement testbed parameters summarized in Table~\ref{tab:system_parameters}. The Tx uses MATLAB-generated Zadoff-Chu (ZC) sequences of length \(128\) for waveform generation, upconverted to the \SIrange{25}{28}{\GHz} via a Sivers EVK02004 RFIC interfaced with a B205mini USRP. The Rx chain performs downconversion and digitization using an identical USRP-RFIC pair. An intermediate frequency (IF) of \SI{3}{\GHz} is used at both the Tx and the Rx. During the measurements, the objects were positioned at the observation point located \SI{3}{\meter} from the Tx-Rx baseline center, with both antennas boresighted (HPBW = \SI{10}{\degree} in elevation/azimuth) to ensure beam convergence at the observation point. Frequency calibration was performed in \SI{1}{\GHz} increments across the \SIrange{25}{28}{\GHz} band, with real-time processing to extract channel impulse responses (CIRs) from the received waveforms for offline processing. All the measurements were conducted in line-of-sight (LoS) propagation conditions with horizontal (H) polarization for both the Tx and the Rx. 
\subsection{Indoor Factory Targets and Measurement Conditions}
\begin{table*}[ht]
\small 
\centering
\caption{Specifications of the InF Objects Used in the Measurements.}
\begin{tabular}{|p{3.7cm}|p{5cm}|p{1.7cm}|p{5cm}|}
\hline
\textbf{Equipment} & \textbf{Dimensions} & \textbf{Mass} & \textbf{Materials} \\ \hline
\textit{Small-Sized UAV}& Folded: \(214\times 91 \times 84\)mm & \(907\)g & Magnesium alloy, reinforced plastic, \\
(DJI Mavic \(2\) Pro)&Unfolded: \(322 \times 242 \times 84\) mm&&carbon fiber, glass, silicon components\\\hline
\textit{Mid-Sized UAV} & Folded: \(430\times 420 \times 430\) mm & \(3.6\)-\(6.3\)kg& Carbon fiber-reinforced plastic, \\ 
(DJI Matrice \(300\) RTK)& Unfolded: \(810 \times 670 \times 430\) mm&&aluminum, plastic \\\hline
\textit{Robotic Arm} & Reach: \(820\) mm & \(\approx 30\)kg & Aluminum, titanium, steel, plastic, \\ 
(KUKA LBR iiwa 14 R\(820\)) &Height: \(1640\) mm (with arm)&& polymer composites\\\hline
\textit{AGV (Quadruped Robot)}  & Length: \(1100\) mm; Width: \(500\) mm; & \(\approx 32.7\)kg & Aluminum, titanium, carbon fiber  \\ 
(Boston Dynamics Spot)& Height: \(610\) mm (walking without arm) &&composites, polymer\\\hline
\end{tabular}
\label{equipment_table}
\end{table*}
\begin{figure}[t]
    \centering
    \subfloat[Small-sized UAV]{
        \includegraphics[width=0.45\columnwidth]{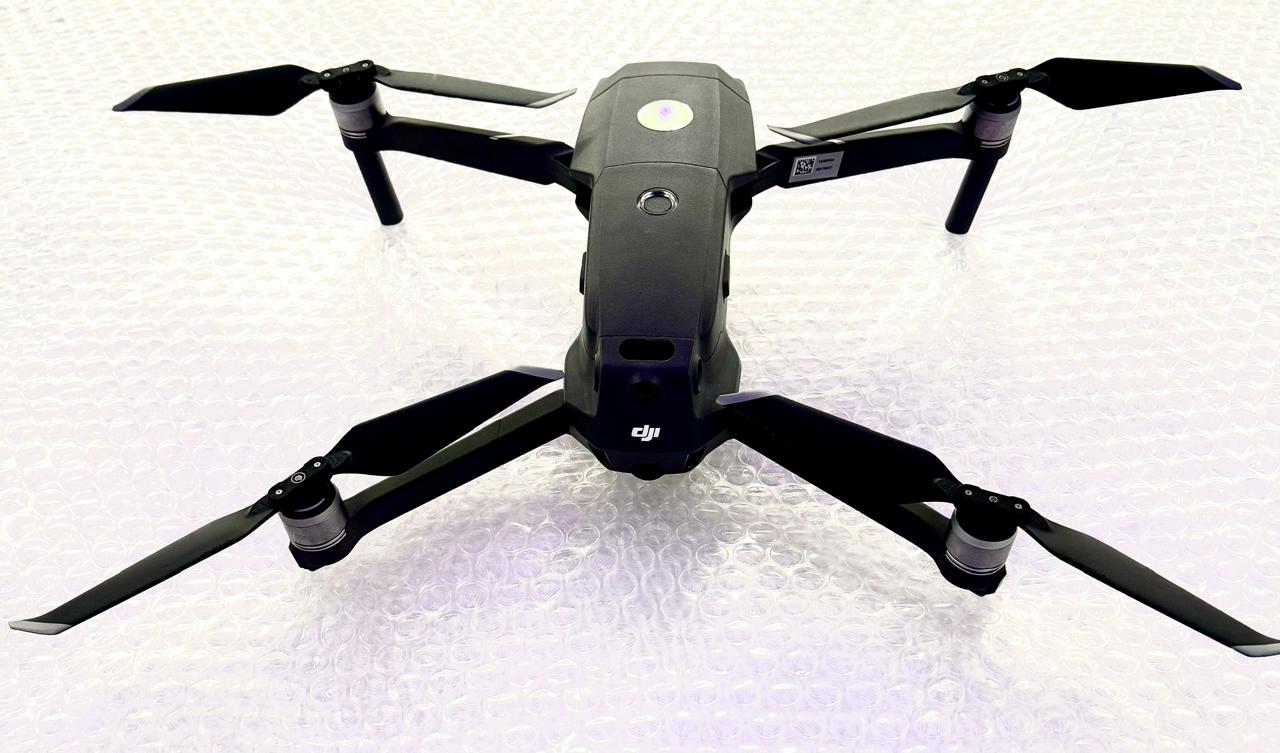}
        \label{drone01}
    }
    \hfill
    \subfloat[Mid-sized UAV]{
        \includegraphics[width=0.43\columnwidth, height=2.4cm, keepaspectratio]{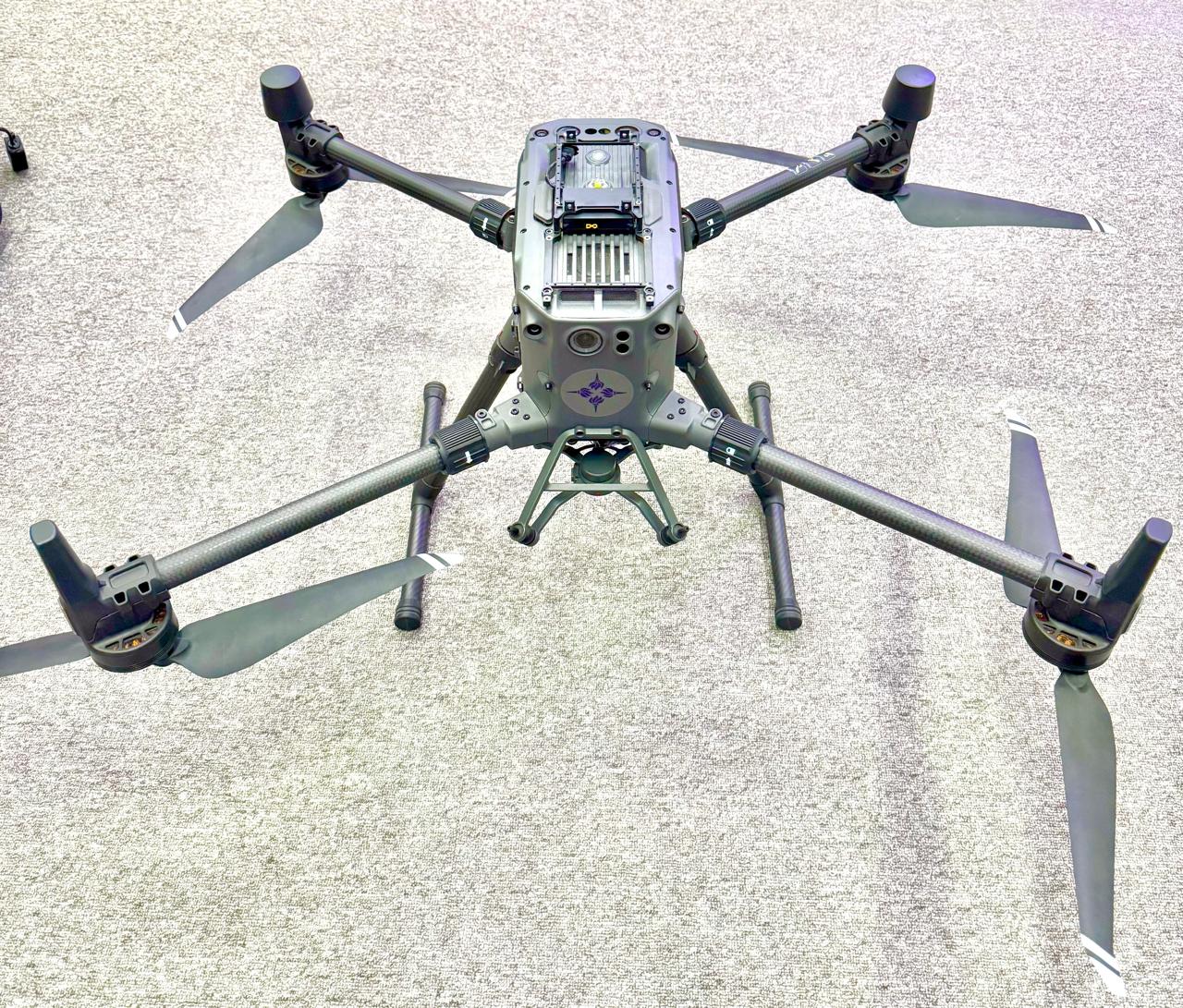}
        \label{drone02}
    }
    \caption{{Small-sized and Mid-sized UAV test objetcs used in the study.}}
    \label{drones}
\end{figure}
\begin{figure}[t]
    \centering
    \subfloat[Robotic Arm]{
        \includegraphics[width=0.8\columnwidth, height=4.6cm, keepaspectratio]{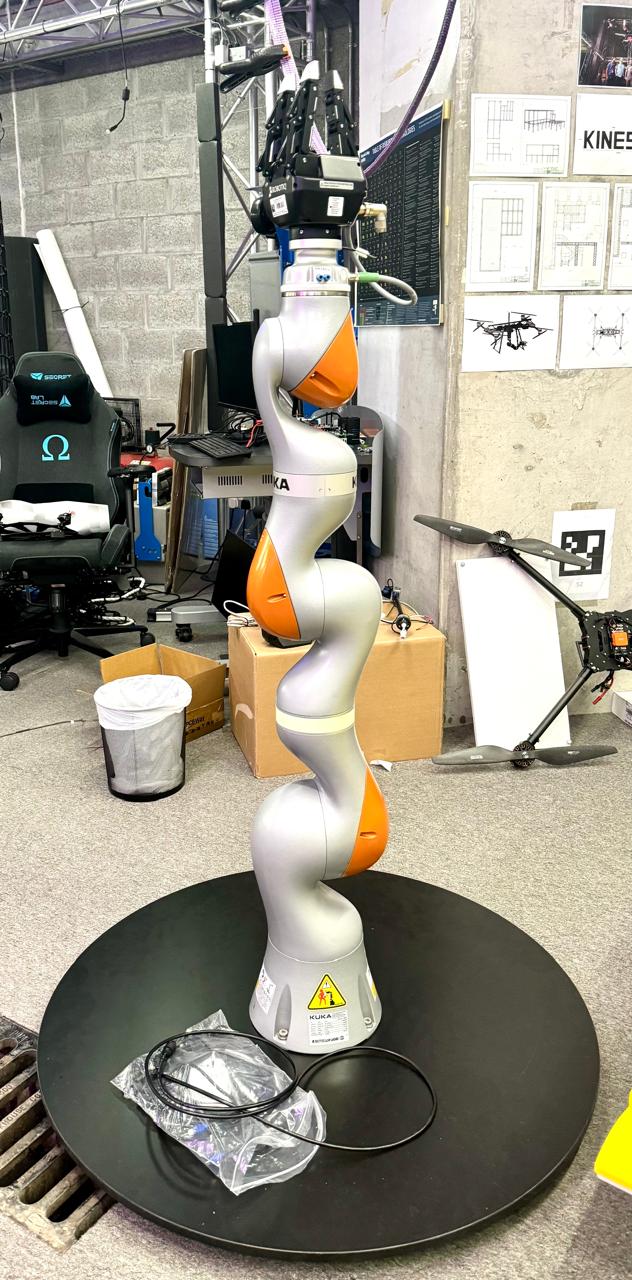}
        \label{robotic_arm}
    }
    \hfill
    \subfloat[AGV]{
        \includegraphics[width=0.43\columnwidth]{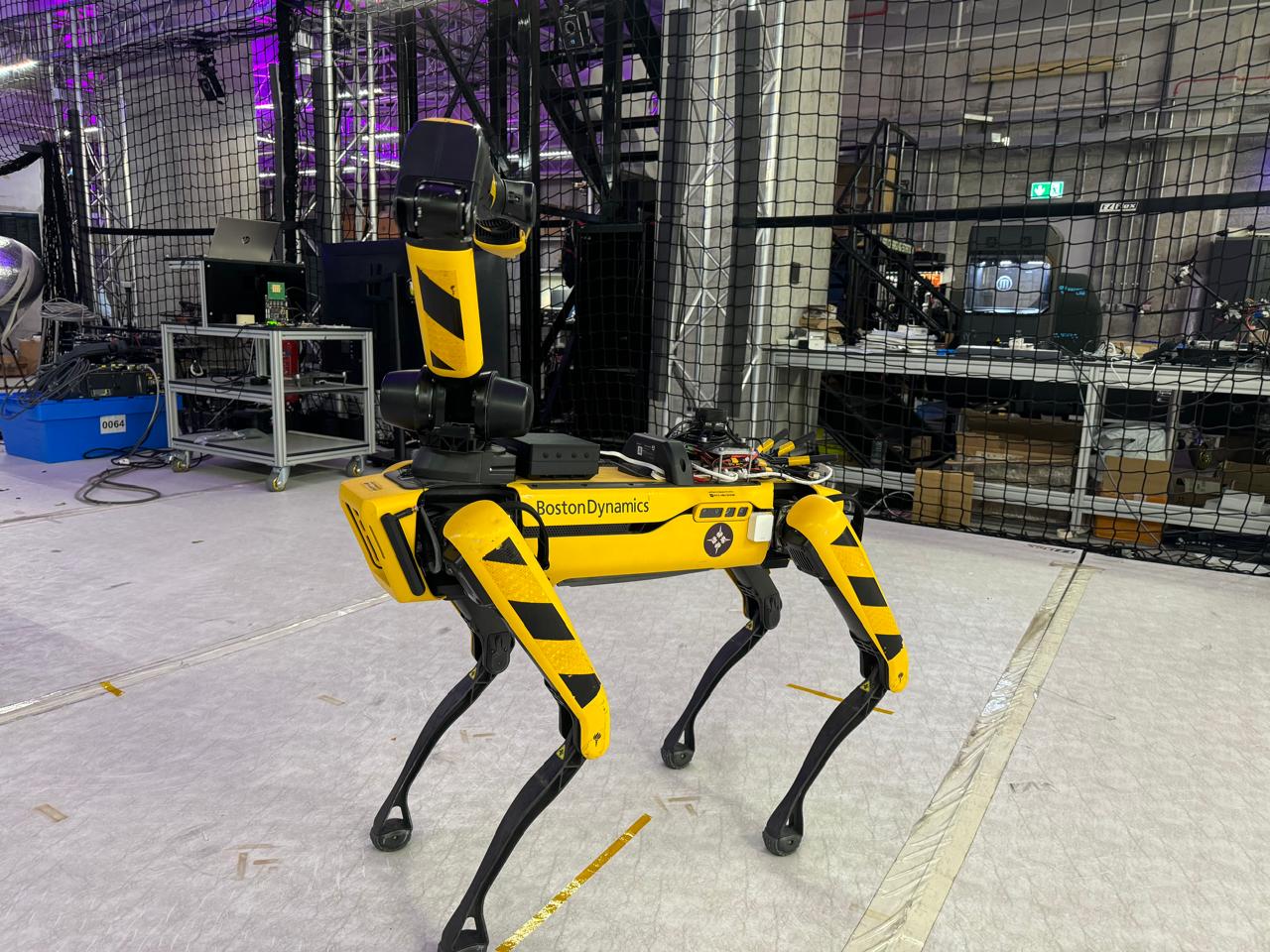}
        \label{robot}
    }
    \caption{Robotic arm and AGV test objetcs used in the study.}
    \label{targets}
\end{figure}
Our measurements focus on objects prevalent in InF environments: (i) small-sized UAV, (ii) mid-sized UAV, (iii) robotic arm, and (iv) AGV, with detailed characteristics provided in Table \ref{equipment_table} and visual representations in Figs. \ref{drones} and \ref{targets}

For UAV measurements, both small and mid-sized platforms were in continuous rotational flight above the observation point, ensuring comprehensive angular coverage of backscatter reflections at the Rx. The vertical clearance between the ground and UAV underside for small-sized UAV was maintained \(0.9\) m and \(0.6\) m for the mid-sized UAV, while laser alignment fixed the geometric center at \(1\) m altitude for both platforms. The mid-sized UAV has dual lithium-ion batteries mounted laterally, whereas, for small-sized UAV has a top-mounted battery. 
\begin{figure}[t]
     \centering
     \subfloat[]{
         \includegraphics[width=0.45\columnwidth, height=5cm, keepaspectratio]{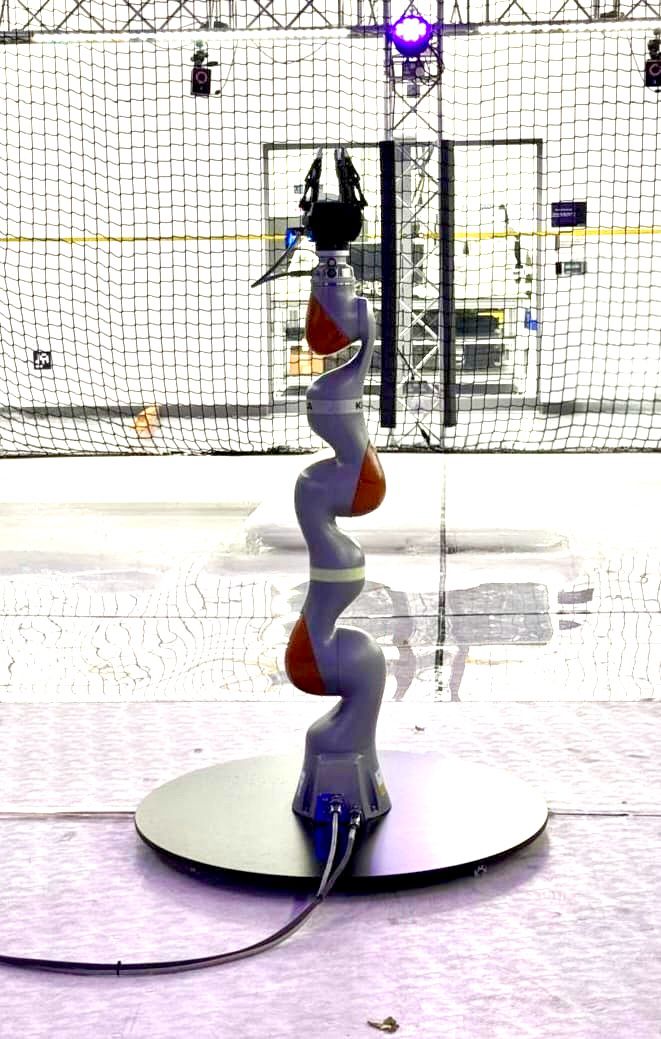}
         \label{ra1}
     }
     \hfill
     \subfloat[]{
         \includegraphics[width=0.43\columnwidth, height=5cm, keepaspectratio]{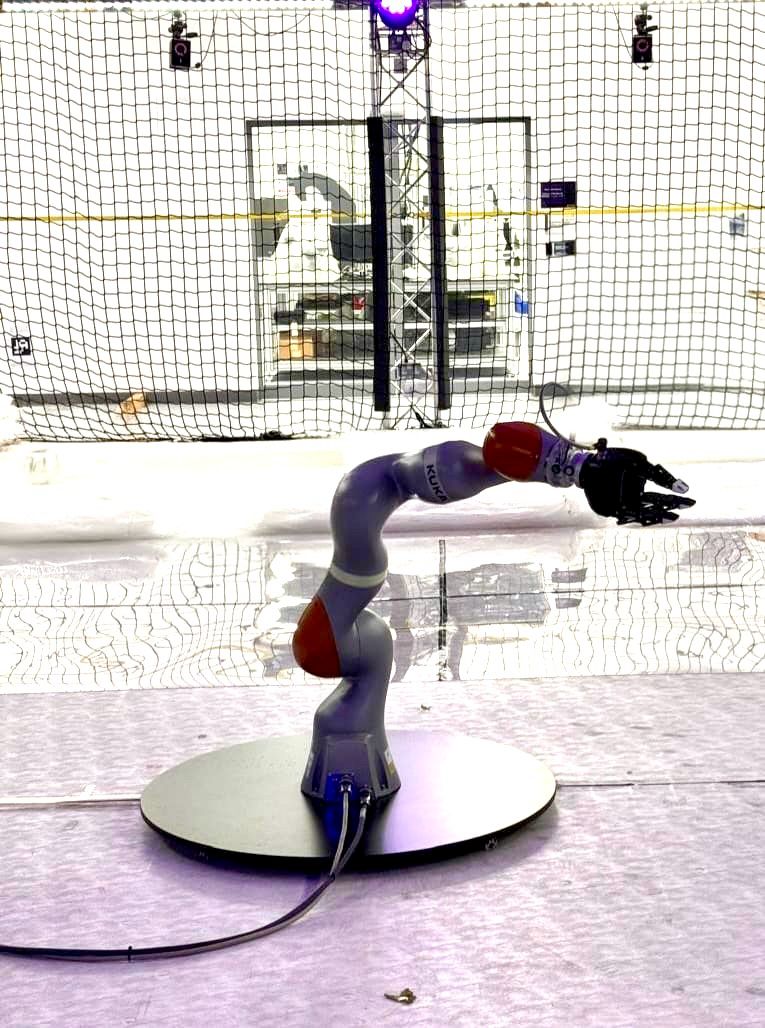}
         \label{ra2}
     }
     \caption{Different motions of robotic arm during the measurements.}
          \label{ra_motions}
\end{figure}

The second object is a robotic arm as shown in Fig. \ref{robotic_arm}. During the measurements, the robotic arm was placed at the observation point while it was executing tasks that are typical to InF operations. The seven joints in the robotic arm enabled continuous motion through typical tasks, including pick-and-place operations, with a couple of them shown in Fig.~\ref{ra_motions}. The robotic arm has a maximum height of \(1640\) mm, however, while emulating different tasks, the height was dynamically reduced during specific motions, resulting in variations in effective reflective surface area. 
 \begin{figure}[t]
     \centering
     \subfloat[]{
         \includegraphics[width=0.45\columnwidth, height=5cm, keepaspectratio]{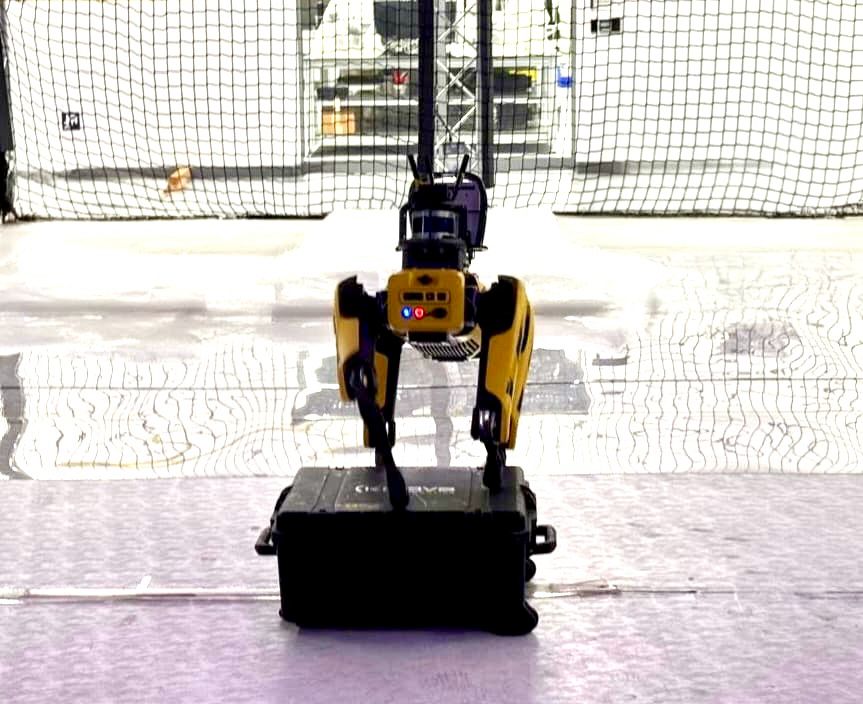}
        \label{ra1}
     }
    \hfill
     \subfloat[]{
         \includegraphics[width=0.43\columnwidth, height=5cm, keepaspectratio]{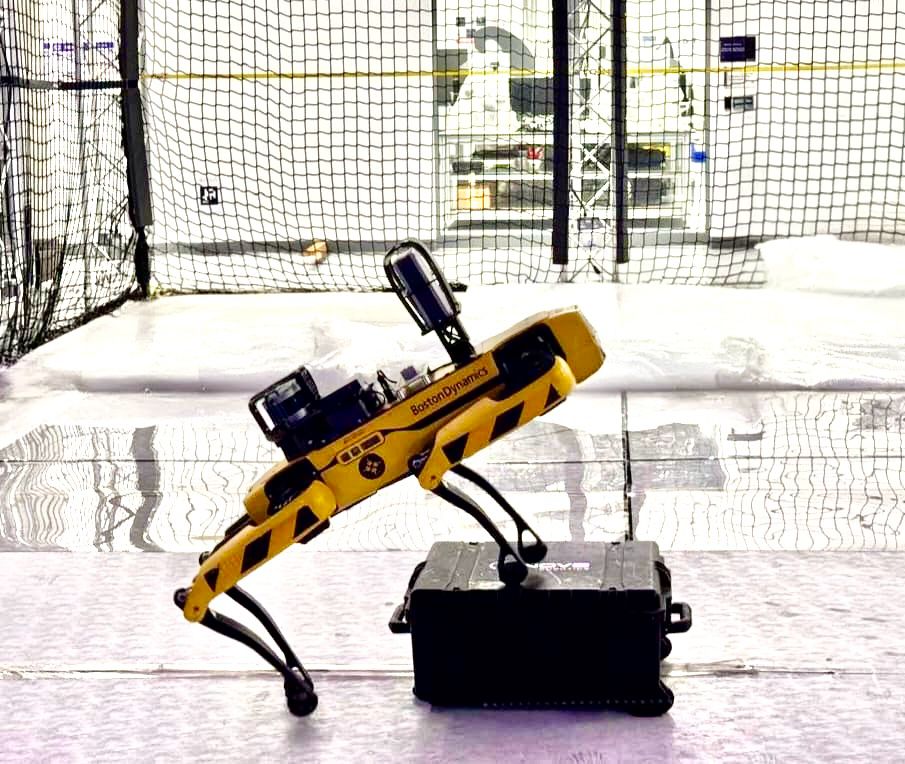}
         \label{ra2}
     }
     \caption{The longitudinal and lateral motion of AGV during the measurements.}
     \label{agv_motions}
 \end{figure}

The third object is an AGV as shown in Fig.~\ref{robot}. During the measurements, two different patterns were analyzed as shown in Fig. \ref{agv_motions}: (i) longitudinal motion toward/away from the observation point for front/back profiles and (ii) lateral motion along the observation point to capture side-profile RCS. Measurements included vertical traversal of a \(20\) cm non-conducting object at the observation point, simulating typical InF obstacle interaction scenarios while maintaining consistent orientation diversity.
\subsection{Experimental Setup and RCS Evaluation Methodology}
\begin{figure}[t]
     \centering
     \includegraphics[trim={0mm 0mm 1mm 0mm},clip, width=0.9\linewidth]{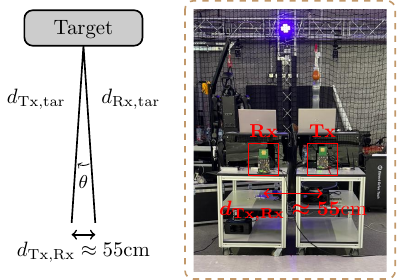} 
     \caption{\textcolor{black}{Quasi-monostatic measurement setup.}}
     \label{setup} 
     \end{figure}

Our experimental setup employs a quasi-monostatic configuration. While ideal monostatic operation requires coincident Tx and Rx positions ($\theta = \SI{0}{\degree}$), practical implementation necessitates spatial separation between the Tx and Rx antennas. The quasi-monostatic measurement configuration as shown in Fig.~\ref{setup} employs a \SI{55}{\centi\meter} Tx-Rx baseline separation $d_\mathrm{Tx,Rx}$, producing an angular offset of $\theta \approx \SI{10}{\degree}$ at the observation point. Both Tx-target $d_\mathrm{Tx,tar}$ and Rx-target $d_\mathrm{Rx,tar}$ distances are maintained at \(\approx\)\SI{3}{\meter}, with the antenna pair centrally positioned in the InF environment as shown in Fig.~\ref{layout}. 

The measurements initiate with acquisition of reference CIR $h_\mathrm{ref}(n)$ for all the operational frequencies, prior to target introduction. The reference measurements characterize the intrinsic contributions of environmental clutter and static scatterers within the measurement domain for the operating frequencies. $h_\mathrm{ref}(n)$ serves as a baseline for subsequent target measurements, and allow to evaluate the reference power, i.e., \(P_\mathrm{ref} = \sum\nolimits_n \vert h_\mathrm{back}(n)\vert^2\). \( P_{\mathrm{ref}} \) quantifies the total energy of the background propagation environment by providing a scalar metric of the ambient clutter power level against which target reflections must be discriminated. Subsequently, the target-inclusive CIR, $h(n)$ is obtained for  all the operating frequencies, capturing the composite effects of environmental scattering $h_{\mathrm{ref}}(n)$ and target-specific reflections $h_{\mathrm{tar}}(n)$. The total received power $P_{\mathrm{tot}} = \sum_n |h(n)|^2$ decomposes into two primary components, that are, $P_{\mathrm{tot}} = P_{\mathrm{tar}} + P_{\mathrm{ref}}$. Through differential power analysis, we isolate the target contribution as $P_{\mathrm{tar}} = P_{\mathrm{tot}} - P_{\mathrm{ref}}$, effectively suppressing environmental clutter. Moreover, it is highlighted that the measurements were performed with high signal power, therefore, even a low variation in received power is above the clutter power and are detectable. 
$P_{\mathrm{tar}}$ computed for all the target object at all the operating frequencies. Since $d_{\mathrm{Tx,tar}} = d_{\mathrm{Rx,tar}} = d$, the target power $P_{\mathrm{tar}}$ directly relates to the RCS $\sigma$ through the following radar equation:
\begin{equation}\label{radar_equation}
P_{\mathrm{tar}} = \frac{P_t G_t G_r \lambda^2 \sigma L}{(4\pi)^3 d^4}
\end{equation}
where $P_t$ is transmit power, $G_t$/$G_r$ are antenna gains, $\lambda$ is the wavelength, and \(L\) are unknown system losses.

The radar equation parameters excluding $\lambda$, $d$, and \(\sigma\) require precise calibration due to their system-specific nature. The calibration process determines a system factor $K(\lambda)$ through measurements where the Rx is positioned at the target location, i.e., observation point, where the procedure leverages the known free-space path loss relationship $P_r = \frac{P_t G_t G_r\lambda^2 L} {(4\pi)^2 d^2}$. Through these measurements, we compute the system factor as $K(\lambda) = P_r/(4\pi d^2)$, by normalizing the received power by $4\pi d^2$ (\(d\) is already know, i.e., \SI{3}{\meter}), which ensures consistency between the single-path (calibration) and dual-path (RCS measurement) scenarios. The calibrated RCS is then derived as $\sigma = K^{-1}(\lambda)P_{\mathrm{tar}}$. The approach ensures compensation for system-specific gains/losses through $K(\lambda)$. The calibration's validity rests on maintaining consistent system parameters (transmit power, antenna patterns, and receiver sensitivity) between calibration and measurement phases.
\section{Results of RCS Fitting to Log-normal Distribution}\label{sec4}
%
\subsection{RCS Fitting to Log-normal Distribution}
The measured RCS data is modeled using a log-normal distribution to capture its asymmetric and heavy-tailed characteristics from combined specular/diffuse reflections. The distribution parameters ($\mu$, $\sigma$) are estimated via maximum likelihood, with goodness-of-fit (GoF) assessed through two complementary metrics. The first one is Kolmogorov–Smirnov (KS) statistic, i.e.,  $\mathrm{KS} = \max |F(x) - F_{\mathcal{LN}}(x)|$ which quantifies the maximum deviation between empirical $F(x)$ and theoretical $F_{\mathcal{LN}}(x)$ cumulative distribution functions (CDFs), and is sensitive to localized mismatches. The second one is the mean squared error (MSE), i.e.,  $\mathrm{MSE} = \frac{1}{N}\sum_{i=1}^N (F(x_i) - F_{\mathcal{LN}}(x_i))^2$ which evaluates global fit quality across all $N$ data points. Small KS/MSE values indicate a better fit. 

\begin{figure}[t]
     \centering
     \includegraphics[trim={2mm 0mm 1mm 0mm},clip, width=1\linewidth]{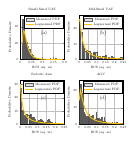} 
     \caption{\textcolor{black}{Empirical and fitted lognormal PDFs for all the InF test objects at \SI{28}{\GHz} considered in the study.}}
     \label{pdfs} 
 \end{figure}

\begin{figure}[t]
     \centering
     \includegraphics[trim={2mm 0mm 1mm 0mm},clip, width=1\linewidth]{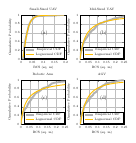} 
     \caption{\textcolor{black}{Empirical and fitted lognormal CDFs for all the InF test objects at \SI{28}{\GHz} considered in the study.}}
     \label{cdfs} 
 \end{figure}

Given space limitations, we present the fitting of RCS data to log-normal distribution by illustrating the probability density function (PDF) and CDF fit for all the objects at \SI{28}{\GHz} in Figs. \ref{pdfs} and \ref{cdfs}. We can observe from Figs. \ref{pdfs} and \ref{cdfs} that log-normal distribution fits the measured RCS data reasonably well as also indicated by the GoF metrics provided in Tables \ref{mavic_params}-\ref{qa_params}. The behavior of the measured RCS data at other frequencies demonstrates consistent trends with variations primarily in parameter values rather than fundamental statistical behavior. The complete set of fitted log-normal parameters ($\mu$, $\sigma$) for all the objects at different operating frequencies is provided in Tables \ref{mavic_params}- \ref{qa_params}.  

\begin{table}[tb]
\centering
\color{black}
\small
\caption{Lognormal Distribution Parameters for Small-Sized UAV.}
\begin{tabular}{|>{\centering\arraybackslash}p{1.7cm}|>{\centering\arraybackslash}p{1.1cm}|>{\centering\arraybackslash}p{1.1cm}|>{\centering\arraybackslash}p{2.9cm}|}
\hline
\textbf{Freq. (GHz)} & \textbf{KS Stat $(\times 10^{-2})$} & \textbf{MSE $(\times 10^{-3})$} & \textbf{Parameters} \\ \hline
\(25\) & \(8.8\) & \(2.3\) & \(\mu = -3.9, \sigma = 1.4\) \\ \hline
\(26\) & \(7.8\) & \(1.4\) & \(\mu = -3.8, \sigma = 0.52\) \\ \hline
\(27\) & \(8.7\) & \(2.7\) & \(\mu = -3.83, \sigma = 1.74\) \\ \hline
\(28\) & \(6\)   & \(0.81\) & \(\mu = -3.79, \sigma = 0.61\) \\ \hline
\end{tabular}
\label{mavic_params}
\end{table}

\begin{table}[tb]
\centering
\color{black}
\small
\caption{Lognormal Distribution Parameters for Mid-Sized UAV.}
\begin{tabular}{|>{\centering\arraybackslash}p{1.7cm}|>{\centering\arraybackslash}p{1.1cm}|>{\centering\arraybackslash}p{1.1cm}|>{\centering\arraybackslash}p{2.9cm}|}
\hline
\textbf{Freq. (GHz)} & \textbf{KS Stat $(\times 10^{-2})$} & \textbf{MSE $(\times 10^{-3})$} & \textbf{Parameters} \\ \hline
\(25\) & \(10\) & \(2.3\) & \(\mu = -3.5, \sigma = 1.42\) \\ \hline
\(26\) & \(7.1\) & \(1.3\) & \(\mu = -3.49, \sigma = 1.47\) \\ \hline
\(27\) & \(14\) & \(6\)   & \(\mu = -3.47, \sigma = 1.96\) \\ \hline
\(28\) & \(8.2\) & \(1.6\) & \(\mu = -3.49, \sigma = 1.48\) \\ \hline
\end{tabular}
\label{rtk_params}
\end{table}

\begin{table}[tb]
\centering
\color{black}
\small
\caption{Lognormal Distribution Parameters for Robotic Arm.}
\begin{tabular}{|>{\centering\arraybackslash}p{1.7cm}|>{\centering\arraybackslash}p{1.1cm}|>{\centering\arraybackslash}p{1.1cm}|>{\centering\arraybackslash}p{2.9cm}|}
\hline
\textbf{Freq. (GHz)} & \textbf{KS Stat $(\times 10^{-2})$} & \textbf{MSE $(\times 10^{-3})$} & \textbf{Parameters} \\ \hline
\(25\) & \(11\) & \(3.2\) & \(\mu = -3.43, \sigma = 1.79\) \\ \hline
\(26\) & \(11\) & \(3.4\) & \(\mu = -3.45, \sigma = 1.83\) \\ \hline
\(27\) & \(8.5\) & \(2.6\) & \(\mu = -3.48, \sigma = 1.82\) \\ \hline
\(28\) & \(13\) & \(5\)   & \(\mu = -3.49, \sigma = 1.67\) \\ \hline
\end{tabular}
\label{ra_params}
\end{table}

\begin{table}[tb]
\centering
\color{black}
\small
\caption{Lognormal Distribution Parameters for AGV.}
\begin{tabular}{|>{\centering\arraybackslash}p{1.7cm}|>{\centering\arraybackslash}p{1.1cm}|>{\centering\arraybackslash}p{1.1cm}|>{\centering\arraybackslash}p{2.9cm}|}
\hline
\textbf{Freq. (GHz)} & \textbf{KS Stat $(\times 10^{-2})$} & \textbf{MSE $(\times 10^{-3})$} & \textbf{Parameters} \\ \hline
\(25\) & \(9.9\) & \(2.1\) & \(\mu = -3.44, \sigma = 1.52\) \\ \hline
\(26\) & \(9.9\) & \(2\)   & \(\mu = -3.44, \sigma = 1.10\) \\ \hline
\(27\) & \(14\)  & \(4.2\) & \(\mu = -3.42, \sigma = 1.3\) \\ \hline
\(28\) & \(8.8\) & \(2\)   & \(\mu = -3.4, \sigma = 1.22\) \\ \hline
\end{tabular}
\label{qa_params}
\end{table}

\begin{figure}[t]
    \centering
    \includegraphics[trim={1mm 0mm 0mm 0mm},clip, width=0.9\linewidth]{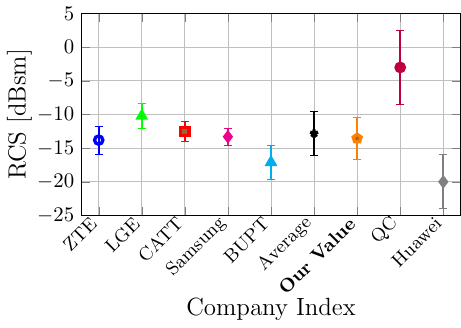} 
    \caption{\textcolor{black}{Monostatic RCS for small-sized UAV (replicated from 3GPP).}}
    \label{comparison} 
\end{figure}
\subsection{\(A\) and \(B_2\) Evaluation}
The RCS follows a log-normal distribution characterized by parameters $\mu$ and $\sigma$, representing the mean and standard deviation of $\ln(X)$ where $X \sim \text{Lognormal}(\mu,\sigma^2)$. The mean RCS in linear scale is given by $\mathbb{E}[X] = e^{\mu + \sigma^2/2}$, with corresponding dBsm value $A = 10\log_{10}(\mathbb{E}[X]) = 10\log_{10}(e^{\mu + \sigma^2/2})$. The normalized variance $B_2$ in linear scale is derived as the squared coefficient of variation $B_2 = \text{Var}(X)/\mathbb{E}[X]^2 = e^{\sigma^2} - 1$, where the variance $\text{Var}(X) = (e^{\sigma^2} - 1)e^{2\mu + \sigma^2}$. When constrained to unit mean RCS ($\mathbb{E}[X] = 1$) as specified by 3GPP, this requires $\mu = -\sigma^2/2$, making $B_2$ dependent solely on $\sigma^2$. The dB-scaled fluctuation parameter becomes $B_2\,(\text{dB}) = 10\log_{10}(e^{\sigma^2} - 1)$, which properly represents relative RCS variations while being consistent with the log-normal framework.
\subsection{Discussion}
The parameters $A$ and $B_2$ are computed for each object at every frequency point following the log-normal distribution analysis. The values are then averaged, with the final consolidated results presented in Table~\ref{tab:RCS_values}. The analysis of Table~\ref{tab:RCS_values} reveals distinct RCS characteristics across different UAV categories. As anticipated, mid-sized UAVs exhibit higher mean RCS values $A$ compared to small-sized UAVs, with this difference attributable to three primary factors: (1) physical dimensions, (2) distinct material compositions, and (3) specular reflections from lithium-ion battery packs. The elevated $B_2$ values observed for mid-sized UAVs further support this interpretation, as rotational dynamics cause alternating strong specular returns when batteries face the radar and diffuse scattering when other surfaces are illuminated. The robotic arm demonstrates particularly interesting RCS behavior, with both high $A$ and $B_2$ values. High \(A\) value is because of its predominantly metallic construction, while the significant $B_2$ emerges from measurement conditions capturing various operational movements. During movement sequences, the effective reflective area varies considerably during the articulation phases. The geometric variability produces the observed wide fluctuation in reflected signal strength, explaining both the high \(A\) and \(B_2\) values. The reduced $A$ value observed for the AGV results from two scattering mechanisms which result in higher number of diffused reflection: (1) the difference between the AGV's operational height (\SI{0.61}{m}) and the Tx-Rx antenna height (\SI{1}{m}), and (2) the longitudinal motion profile presents an effective reflective area that is much smaller than lateral reflective area.

The derived log-normal parameters for small-sized UAVs yield $A = \SI{-13.57}{dBsm}$ and $B_2 = \SI{3.065}{dB}$, demonstrating close agreement with 3GPP's standardized values of $A = \SI{-12.81}{dBsm}$ and $B_2 = \SI{3.74}{dB}$ established through industry consensus. The marginal deviations ($\Delta A = \SI{0.76}{dB}$, $\Delta B_2 = \SI{0.675}{dB}$) fall well within the expected tolerance for RCS characterization, as evidenced from different industrial proposals. To further validate this agreement, we replicate the 3GPP’s monostatic RCS distribution graph for small-sized UAVs, superimposing our values as shown in Fig. \ref{comparison}. The near-overlap of the curves particularly in the \SIrange{-15}{-10}{dBsm} range confirms accuracy of our RCS characterization approach with the 3GPP standardized values. 

\begin{table}[tb]
\centering
\small
\caption{RCS values for the given targets considering 3GPP framework.}
\label{tab:RCS_values}
\begin{tabular}{|>{\centering\arraybackslash}p{2.3cm}|>{\centering\arraybackslash}p{1.5cm}|>{\centering\arraybackslash}p{1.5cm}|>{\centering\arraybackslash}p{1.5cm}|}
\hline
\textbf{Target} & \({A}\) (dBsm) & \({B_1}\) (dB) & \({B_2}\) (dB) \\
\hline
Small-Sized UAV& \(-13.57\) & \(0\) & \(3.065\) \\ \hline
Mid-Sized UAV & \(-9.6\) & \(0\) & \(10.66\) \\ \hline
Robotic Arm & \(-8.165\) & \(0\) & \(13.54\) \\ \hline
AGV & \(-11.235\) & \(0\) & \(6.27\) \\ \hline
\end{tabular}
\end{table}
\section{Conclusions}\label{sec5}
The experimental results provided in this paper establish three principal findings for InF RCS characterization: First, small-sized UAV measurements confirm 3GPP's standardized parameters within \SI{1}{dB} tolerance, validating our evaluation methodology. Second, mid-sized targets exhibit \SIrange{3}{4}{dB} higher reflectivity than small UAVs, attributable to their larger physical dimensions and prominent battery reflections. Robotic arm demonstrate unique RCS signatures with \SI{13.54}{dB} fluctuation variance during operation, directly correlated with mechanical articulation patterns. The AGV measurements further reveal how target object height and motion profiles affect scattering characteristics. Our findings validate our methodology for RCS characterization against the standardized 3GPP characterization approach.
\bibliographystyle{unsrt}
\bibliography{main}

\end{document}